\documentclass{article}
\usepackage{ijcai21}

\usepackage{soul}
\usepackage{url}
\usepackage[hidelinks]{hyperref}
\usepackage[utf8]{inputenc}
\usepackage[small]{caption}
\usepackage{graphicx}
\usepackage{amsmath}
\usepackage{amsthm}
\usepackage{booktabs}
\usepackage{algorithm}
\usepackage{algorithmic}
\usepackage{balance}
\usepackage{multirow}
\usepackage{latexsym}

\newcommand{\comment}[1]{}

\title{Sequence-to-Sequence Learning on Keywords \\ for Efficient FAQ Retrieval\footnote{\it Published at the IJCAI 2021 Workshop on Applied Semantics Extraction and Analytics (ASEA).}}

\author{
Sourav Dutta
\and
Haytham Assem \and
Edward Burgin \\
\affiliations
Huawei Research Centre, Dublin, Ireland \\
\emails
\{sourav.dutta2, haytham.assem, edwardburgin\}@huawei.com,
}

\begin{document}
\maketitle

\begin{abstract}
	Frequently-Asked-Question (FAQ) retrieval provides an effective procedure 
	for responding to user's natural language based queries. Such platforms 
	are becoming common in enterprise chatbots, product question answering, 
	and preliminary technical support for customers. However, the challenge 
	in such scenarios lies in bridging the {\em lexical and semantic gap} 
	between varied query formulations and the corresponding answers, both 
	of which typically have a {\em very short span}.

	This paper proposes {\em TI-S2S}, a novel learning framework combining
	TF-IDF based keyword extraction and Word2Vec embeddings for training a 
	Sequence-to-Sequence (Seq2Seq) architecture. It achieves high precision 
	for FAQ retrieval by better understanding the underlying {\em intent} 
	of a user question captured via the representative keywords. We further 
	propose a variant with an additional neural network module for guiding 
	retrieval via relevant candidate identification based on similarity features. 
	Experiments on publicly available dataset depict our approaches to 
	provide around $92\%$ precision-at-rank-$5$, exhibiting
	nearly $13\%$ improvement over existing approaches.
\end{abstract}

\section{Introduction}
\label{sec:intro}

Frequently-Asked-Questions (FAQ) provide a collection of question-answer pairs that are either manually 
created or automatically extracted from relevant documents. FAQ provide users with an ``one-stop''
source for the most relevant or most searched information pertaining to a product or service -- to enable 
prompt customer help for general queries.

\vspace*{1mm}
{\bf Motivation.} FAQ retrieval systems provide a natural language interface for querying an FAQ collection, 
and is thus increasingly becoming popular with large-scale service-providing companies for presenting information 
to customers. Such systems provide two-fold advantages: (i) automation of customer service tasks, e.g., intelligent 
chatbots~\cite{massaro2018,yan2016} and automated e-mail answering~\cite{karan2018,malik2007}, and (ii) enable 
efficient access to internal FAQ documents for customer service agents, increasing the quality and efficiency.
Further, with the advent of personal assistants (like XiaoIce, Siri, Alexa, Google Assistant, etc.),
these ``virtual agents'' can provide answers to questions and help users solve routine tasks by an additional channel 
to FAQs, hotlines, and forums -- enabling a natural interaction with users~\cite{lommatzsch2019,por2020}.

\vspace*{1mm}
{\bf Challenges.} 
FAQ retrieval is a challenging task, majorly attributed to the fact that the question-answer texts are short, making it 
harder to bridge the {\em lexical and semantic gap} between a user query and the FAQ questions due to short span with 
limited context~\cite{karan2018,lee2008}. Further, precise understanding of user questions can be difficult due to 
informal representations, domain-specificity, abbreviations, and formal-colloquial term mismatches~\cite{lommatzsch2019}. 
For example, consider the questions {\small{\tt ``How can I seal a hole in the gas tank of 
my car?''}} and {\small{\tt ``How to patch a leak in the fuel compartment of my van?''}} which are semantically matching but 
exhibit low lexical overlap and formal-colloquial mismatch.
In addition, FAQ retrieval systems should be able to handle both keyword as well as {\em short span} ``natural language'' questions.
Given the predominantly ``customer-centric'' nature, such systems generally demand higher precision and interpretability 
compared to traditional information retrieval methods.

\vspace*{1mm}
{\bf Problem Statement.} The task of {\em FAQ Retrieval} entails the efficient ranking (in terms of relevance) of 
question-answer pairs of a collection, in response to a user input query. In other words, such retrieval engines 
attempt to understand the underlying {\em intent} of users and retrieve the most related answers 
containing the correct information~\cite{kothari2009}. 

Formally, consider a pre-curated collection (or repository) of question-answer $(Q, A)$ pairs to comprise the FAQ $= \{(Q_1, A_1), \cdots, (Q_n, A_n)\}$, 
where $Q_i$ denotes a question related to the domain, and $A_i$ represents the corresponding answer. Given a user query $q$, 
the task then is to return an ordered list of FAQ $(Q, A)$ pairs, $\{(Q_1^q, A_1^q), \cdots, (Q_k^q, A_k^q)\}$, depicting high semantic and intent similarity 
with respect to the input user query $q$.

\vspace*{1mm}
{\bf Contributions.} This work proposes {\em TI-S2S}, a novel keyword based supervised learning framework for efficient FAQ retrieval. 
Our approach leverages {\em sequence-to-sequence} model to generate representative labels for user questions to compute query-question 
similarity. Additionally, a variant incorporating ``candidate'' identified using a deep learning architecture to guide the retrieval process 
is shown to further improve performance. 

We show that our proposed framework efficiently captures: (i) domain-specificity of the application, (ii) characteristic words and / or concepts 
to differentiate between questions, and (iii) semantic similarity for retrieving relevant QA pairs from the FAQ collection.

Experiments on public FAQ dataset depict our framework to outperform existing techniques in terms of accuracy, 
and also in robustness to limited training data.
In effect, it implicitly considers both {\em document redundancy} and {\em query redundancy}~\cite{karan2018}.

\section{Related Work}
\label{sec:rel}

The problem of FAQ retrieval lies at the intersection of information retrieval and question answering and have thus been studied using 
techniques from both the fields. Initial works on FAQ retrieval relied on manual feature engineering based on 
text similarities using parsing, edit distance, TF-IDF measures, longest common subsequence~\cite{kothari2009}, match-template 
construction~\cite{sneider2010}, and statistical approaches~\cite{berger2000} to name a few. The use of both query-question and 
query-answer vector space similarities within a ranking model was studied in~\cite{jij2005}. However observe, over-emphasis on 
query-answer similarity would be inefficient in scenarios where significant parts of different answers might be similar. 
For example, answers to both the questions {\small{\tt ``How to add an account photo?''}} and {\small{\tt ``How to change the account name?''}} 
might possibly share the common snippet {\small{\tt ``Go to Account > Setting > Profile ...''}} or similar. Further, answers might 
change depending on updates to processes and manuals which might necessitate costly re-training of the entire framework. Such scenarios 
might degrade the performance of approaches based on query-answer similarities~\cite{qexp18,sakata2019}. Thus, in our setting, we do not 
consider the answer to form a part of the retrieval process.

Contextualized language models like BERT~\cite{devlin2019} have been shown to capture semantic relatedness, 
and such embedding techniques have been coupled with traditional IR techniques for FAQ Retrieval~\cite{sakata2019}. 
The use of knowledge graphs have also been studied for Question-Answering (Q-A), by use of entity-concept ``anchors'' in this context~\cite{xie2019}.
Deep Learning has recently enjoyed significant success in classification tasks by constructing high-dimensional latent 
feature space. 
A neural network with word embeddings was proposed in~\cite{yan2016}, while a convolution 
neural network (CNN) based learning-to-rank module was presented in~\cite{karan2018}. However, supervised methods require large 
FAQ-collection with annotations, which are expensive. Hence, in practice such annotated datasets are usually too small to meaningfully train 
complicated machine learning models. Further, such models tend to face difficulty in handling long-tailed questions.
To tackle the problem of limited context in FAQ systems attention-based deep learning models~\cite{gupta19}, query expansion~\cite{qexp18} and query generation~\cite{mass20} 
have recently been studied. Document ranking via sequence-to-sequence has also been studied~\cite{nog20}.

Community and non-factoid question answering (CQA)~\cite{nonfact2011,cqa2017} are closely related, but involve larger corpus with 
broader scope, and hence is not directly applicable to FAQ Retrieval, typically with context brevity and smaller training data.

\section{{\em TI-S2S} Framework}
\label{sec:tis2s}

This section describes the working of our proposed {\em TF-IDF Induced Sequence-to-Sequence} (TI-S2S) algorithm for efficient 
FAQ Retrieval. It couples TF-IDF score to extract keywords (modeling intents in user queries) and word embeddings (capturing 
semantic similarity among questions) for learning a {\em sequence-to-sequence} model to transform syntactically 
different but semantically similar questions into a {\em common representative sequence}. 

Given an FAQ collection (set of question-answer (QA) pairs), our {\em TI-S2S} framework hinges on the following:

\vspace*{1mm}
{\bf A. Pre-Processing.} The questions in the input FAQ collection are initially pre-processed to remove stopwords and are lemmatized. 
For each question $Q_i$, several variations of the question are created (either manually or by automated paraphrasing techniques) or 
are extracted (from query logs via duplicate detection or similarity measures). Such semantically similar paraphrased questions are 
added to the FAQ and are annotated to depict that they convey the same user information intent. As proposed in~\cite{karan2018}, the 
paraphrased QA pairs in FAQ along with the relevance annotations are used for supervised training. Index structures storing the relevance 
information between questions are constructed to assist the subsequent modules.

\vspace*{1mm}
{\bf B. Intent Target Keyword Learning.} 
Based on the relevance annotations among the questions in the FAQ, {\em TI-S2S} creates groups or clusters of questions that are semantically 
similar to (or paraphrases of) each other. For each group of such similar questions (or annotated paraphrased variants) in the FAQ, we extract 
words that have TF-IDF score~\cite{tfidf} (computed on the entire FAQ collection) greater than a thresholding parameter $\tau$, denoted as 
{\em intent target keywords}. Intuitively, these intent keywords capture the context and topic of the question groups.
%, and can be logically considered to be the important keywords modeling the topics of the questions. 
Hence, these {\em intent keywords} enable a ``common representative sequence'' for each group of similar questions (refer Table~\ref{tab:tfidf} for example), 
providing cues for {\em weak supervision} in training the subsequent modules.

\vspace*{1mm}
{\bf C. Seq2Seq Learning.} A sequence-to-sequence (Seq2Seq) model~\cite{seq2seq} utilizes an encoder-decoder architecture for learning to transform 
an input sequence to a corresponding output sequence (possibly of differing lengths).
% and have been shown to perform extremely well for machine translation applications. 
{\em TI-S2S} uses a Seq2Seq module to {\em learn to transform} a question $Q_i \in FAQ$ (i.e., a sequence of pre-processed words) to the {\em 
representative intent target keyword sequence} associated with the question group to which $Q_i$ belongs to. Word embeddings of the questions 
(using Word2Vec~\cite{mikolov2013}) are fed to the input layer of the Seq2Seq module for training with 
%for it to learn semantic similarities between related questions. We 
{\em teacher-forcing} technique~\cite{bengio2015} and Luong attention mechanism~\cite{luong2015}.

It is interesting to note that the transformation of questions into a common keyword space bridges the lexical gap, while the use of word embeddings 
(of the question) bridges the semantic gap. For example, both the words `image' and `photo' (similar in the embedding space) in different questions 
would be trained to generate the same output word `picture' (a common representative keyword) from the seq2seq module -- addressing the lexical and 
semantic gap between user and FAQ questions with short spans.

\vspace*{1mm}
{\bf D. Translated FAQ.} The above trained Seq2Seq model is then used to transform the questions in FAQ to {\em intent representative 
format}. That is, this module translates the input FAQ into a collection of 3-tuples $\{(Q_i, \overline{Q}_i, A_i)\}$ -- where $(Q_i, A_i) \in$ FAQ is 
the original QA pair and $\overline{Q}_i$ is the {\em predicted intent keyword sequence} for $Q_i$ obtained from the Seq2Seq module.

Ideally, the predicted $\overline{Q}_i$ should be the same as the intent target keywords (provided during training) associated to the 
question group to which $Q_i$ (and other similar or paraphrased questions) belongs to. However, in practice, training losses and presence of noise might 
lead to deviations. Through this, {\em TI-S2S} aims to minimize the impact of such error propagation to the final phase.

\vspace*{1mm}
{\bf E. FAQ Retrieval.} The trained {\em TI-S2S} framework along with the translated FAQ forms the proposed FAQ Retrieval platform for user queries. 
Given a new user query $q$, it is initially pre-processed and its word embeddings (as in {\em Modules A} and {\em C} above) are provided as input to 
{\em TI-S2S}.
The ``predicted intent target keyword sequence'' ($\overline{q}$) from the Seq2Seq module is then compared with all $\overline{Q}_i$ in the translated FAQ. 
A similarity score between $\overline{q}$ and $\overline{Q}_i$ is used to obtain the final ranked list of QA pairs of the FAQ. To capture syntactic and 
semantic similarity between the keyword sequences, we use the average of Word Mover's Distance~\cite{wmd} and Levenstein distance between $\overline{q}$ and $\overline{Q}_i$.

Since, the final stage uses Word Mover's Distance and Levenstein distance to compute the similarities between the representative sequences (treated as 
bag-of-significant-words), the order of the predicted representative sequence (obtained from the Seq2Seq module) is not important and our framework is not 
sensitive to it. This provides flexibility to our framework and does not enforce strict order in the seq2seq generation process. Further, the modular structure 
of our framework enables it to be easily adapted to diverse application scenarios with algorithmic variants -- attention mechanisms for Seq2Seq learning or 
combinations of different similarity measures.

\subsection{{\em GTI-S2S} Variant}
\label{ssec:gtis2s}

We now present {\em Guided TF-IDF Induced Sequence-to-Sequence} (GTI-S2S), a variant of the {\em TI-S2S} framework to cater to scenarios with high domain-specificity 
and noisy training process. {\em GTI-S2S} (along with Seq2Seq module) employs an additional recurrent neural network (RNN) to learn to predict {\em question-question relevance} 
using features like entity overlap, Levenstein distance and embedding space similarity between two input questions. 

Thus, given the groups of similar or paraphrased questions (as discussed in {\em Module A}), the RNN is trained as a binary classifier to predict if two 
questions are similar and / or relevant (using the relevance annotations), thus providing {\em ``guided candidate QA selection''} during the retrieval phase of {\em TI-S2S} framework 
(Section~\ref{sec:tis2s}).

Specifically, on arrival of a user query $q$, the predicted intent target keyword sequence ($\overline{q}$) are generated by {\em TI-S2S} (as in {\em Module E}). Additionally, 
for each question $Q_i \in$ FAQ, {\em GTI-S2S} now computes its relevance to $q$ (using the above trained RNN module). Based on the predicted classification probabilities, the 
top-$k$ FAQ questions (with probabilities above a threshold) are extracted as ``prime candidates'' for the user query. Finally, the similarity scores between the obtained 
candidate questions' predicted keyword sequence and $\overline{q}$ are computed to obtain the final rank list. 

In a nutshell, {\em GTI-S2S} can be viewed as a two-stage framework: (a) generation of candidates using RNN and (b) use of TI-S2S framework for ranking the candidates. While in the TI-S2S 
framework, the final similarity of the user question (after prediction phase using seq2seq) is computed against all the questions in the FAQ collection; in GTI-S2S, the final similarity 
is computed only with the candidates identified from the stage (a). Later in Section~\ref{ssec:result} we show the performance advantages of ``candidate generation guided retrieval'' in 
{\em GTI-S2S} in certain settings.

\section{Experimental Results}
\label{sec:expt}

%\comment{
{\small{
\begin{table*}[t]
	\centering
	\caption{(a) Performance of algorithms on {\em MAP} and {\em P@5} measures. (b) Effect of TF-IDF threshold ($\tau$) on MAP and P@5.}
	\label{tab:all_results}
	\begin{tabular}{||c|cc||cc}
		\multicolumn{3}{c}{(a)} & \multicolumn{2}{c}{\hspace*{9mm}(b)} \\
		\cline{1-3}
		{\bf Approaches} & {\bf MAP} & {\bf P@5} & & \multirow{8}{*}{\hspace*{2mm}\includegraphics[width=3in, height=1.5in]{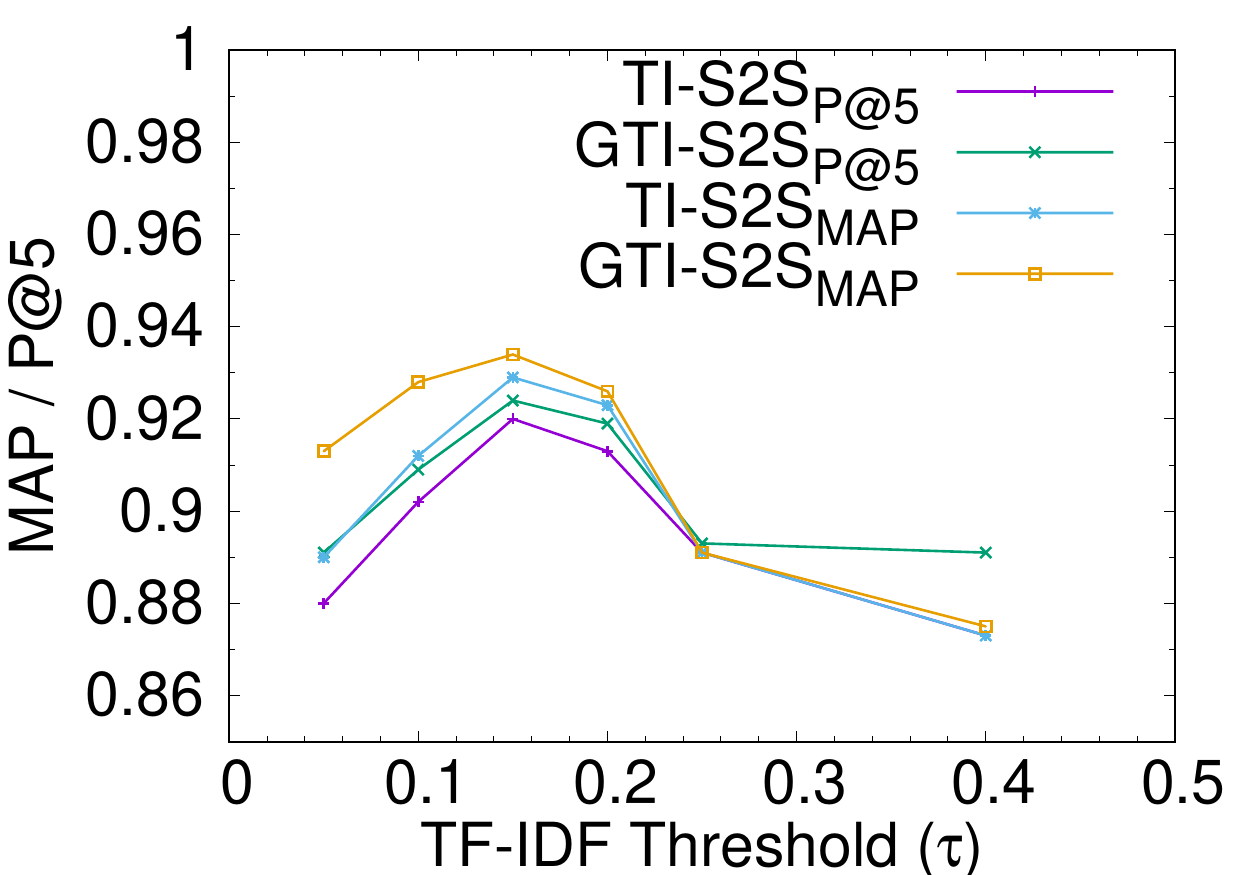}}\\
		\cline{1-3}
		\cline{1-3}
		{\em CNN-Rank}~\cite{karan2018} & 0.74 & 0.62 & & \\
		{\em TSU-BERT}~\cite{sakata2019} & 0.897 & 0.776 & &\\
		{\em BERT}~\cite{devlin2019} & 0.614 & 0.583 & & \\
		{\em RoBERTa}~\cite{liu2019} & 0.712 & 0.796 & & \\
		{\em SBERT}~\cite{reimers2019} & 0.686 & 0.774 & & \\
		{\em TI-S2S} & {\it 0.929} & {\it 0.92} & & \\
		{\em GTI-S2S} & {\bf 0.934} & {\bf 0.924} & & \\
		\cline{1-3}
		\multicolumn{5}{c}{} \\
	\end{tabular}
\end{table*}
}}
%}

\comment{
\begin{table*}[t]
	\centering
	\begin{tabular}{||c|cc||}
		\toprule
		{\bf Approaches} & {\bf MAP} & {\bf P@5} \\
		\midrule
		\midrule
		{\em CNN-Rank}~\cite{karan2018} & 0.74 & 0.62 \\
		{\em TSU-BERT}~\cite{sakata2019} & 0.897 & 0.776 \\
		{\em BERT}~\cite{devlin2019} & 0.614 & 0.583 \\
		{\em RoBERTa}~\cite{liu2019} & 0.712 & 0.796 \\
		{\em SBERT}~\cite{reimers2019} & 0.686 & 0.774 \\
		{\em TI-S2S} & {\it 0.929} & {\it 0.92} \\
		{\em GTI-S2S} & {\bf 0.934} & {\bf 0.924} \\
		\bottomrule
	\end{tabular}
	\caption{Performance of algorithms on {\em MAP} and {\em P@5} measures. {\small (Note: Bold numbers represent the best results, while results in italics denote the second best.)}}
	\label{tab:all_results}
\end{table*}

\begin{figure}[t]
	\centering
	\includegraphics[width=\columnwidth]{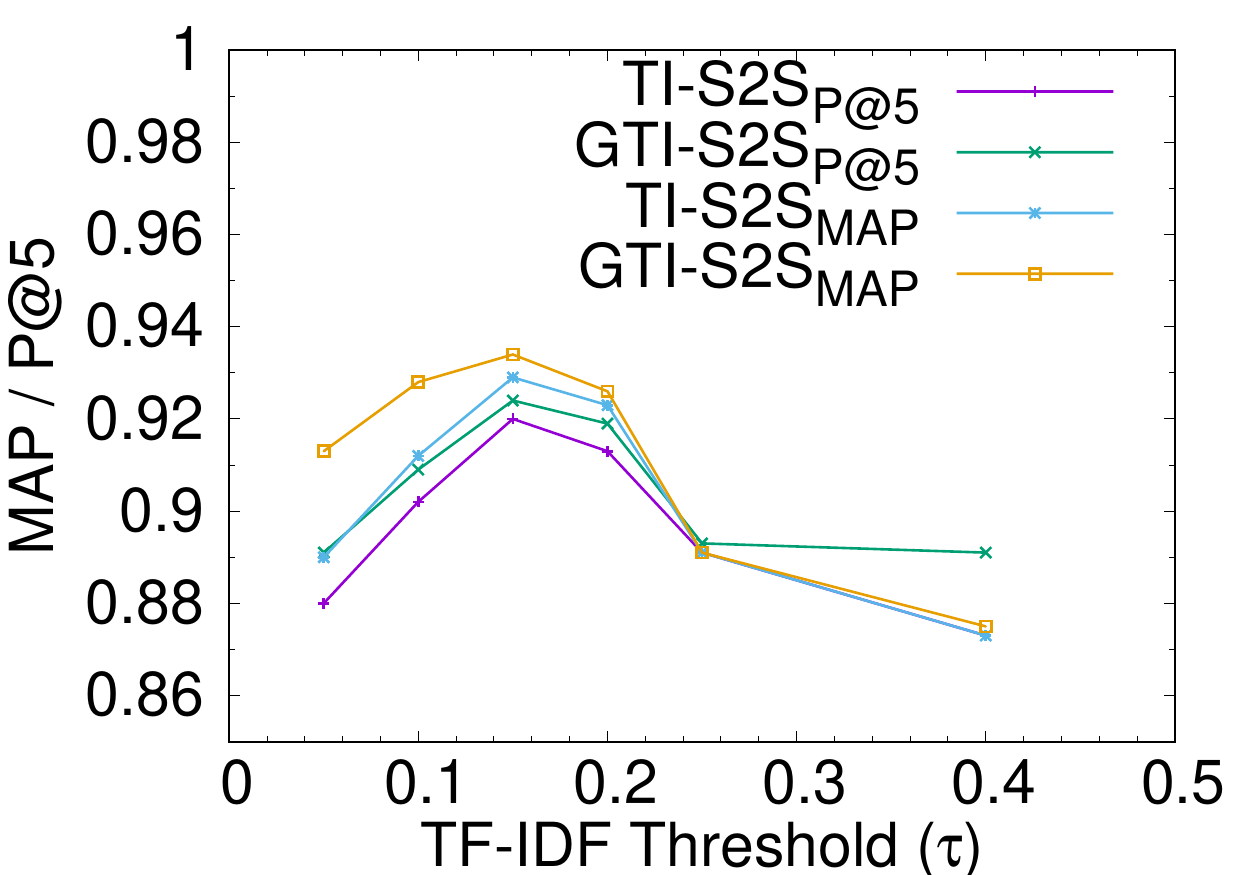}
	\caption{Effect of TF-IDF threshold ($\tau$) on MAP and P@5 performance.}
	\label{fig:tfidf}
\end{figure}
}

\begin{table*}[t]
	\centering
	\vspace*{2mm}
	\hspace*{-5mm}
	\caption{Representative intent keywords extracted for different question clusters on StackExchange data.}
	\label{tab:tfidf}
	\resizebox{2.2\columnwidth}{!}{%
	\begin{tabular}{|c|cccc|}
		\hline
		{\bf Sample Questions} & ${\bf \tau \geq 0.4}$ & ${\bf \tau \geq 0.25}$ & ${\bf \tau \geq 0.15}$ & ${\bf \tau \geq 0.05}$ \\
		\hline
		How secure is my sensitive data on dropbox & & & attacker; data; & attacker; concern; data; dropbox; \\
		Are there security threats to dropbox & & dropbox; & dropbox; secure; & eavesdrop; file; good; information; \\
		Is sensitive data on dropbox secure & dropbox; & security; & security; sensitive; & know; like; malicious; safe; secure; \\
		Does dropbox have good security against attackers & security; & threat; & threat; & steal; threat; tight; transfer; \\
		How safe is my data on dropbox & & & & sensitive; use; user; \\
		\hline
		Is splitting conversations possible in mail threads on gmail & & conversation; & gmail; manage; & assign; bcc; break; confusing; divide; \\
		Can i split a conversation in gmail & & gmail; & conversation; one; & conversation; easy; gmail; hard; keep; \\
		How do I split two merged gmail conversations & conversation; & split; & merge; separate; & large; mail; manage; merge; people; \\
		Splitting conversations in gmail & & thread; & split; thread; & one; possible; reply; response; small; \\
		How to divide a conversation on gmail & & & & separate; split; thread; time; track; \\
		\hline
	\end{tabular}
}
\end{table*}

We now compare the performance of our proposed framework with competing state-of-the-art approaches for FAQ retrieval on open dataset.

\vspace*{1mm}
{\bf Dataset Used.} We perform experiments on the publicly available {\em StackExchange} FAQ dataset~\cite{karan2018} (from {\scriptsize \url{www.takelab.fer.hr/data/StackFAQ/}}). 
It contains $125$ QA threads pertaining to popular Web applications, with each thread containing an original query and $10$ different manual paraphrasings (annotated as relevant to the original question) 
-- a total of $1375$ $(Q, A)$ pairs. The task is then to return a ranked result of the QA pairs in terms of their relevance to a query, as in the setup of~\cite{karan2018,sakata2019}.
%, binary relevance judgments for all ``train'' queries were used for supervised learning.

{\bf Competing Approaches.} We benchmark the performance of our proposed framework against the following methods:\\
(1) {\em CNN-Rank}~\cite{karan2018} -- uses learning-to-rank via convolutional NN architecture. \\
(2) {\em TSU-BERT}~\cite{sakata2019} -- combination of TSUBAKI IR engine for computing query-question and BERT based embeddings for query-answer similarities (from {\small \url{github.com/ku-nlp/bert-based-faqir}}).\\
(3) {\em BERT}~\cite{devlin2019} -- bidirectional language representation fine-tuned to capture contextual similarities using cosine score (from {\small \url{github.com/hanxiao/bert-as-service}}). \\
(4) {\em RoBERTa}~\cite{liu2019} -- fine-tuned optimized version of BERT for better contextual similarity computation (from {\small \url{github.com/pytorch/fairseq/blob/master/examples/roberta}}). \\
(5) {\em SBERT}~\cite{reimers2019} -- Siamese network structure for sentence embeddings using {\em roberta-large-nli-stsb-mean-tokens}), particularly suitable for FAQ retrieval given the short span of texts 
(from {\small \url{github.com/UKPLab/sentence-transformers}}.

{\bf Fine-Tuning.} The BERT and RoBERTa baselines were fine-tuned on the training data to identify similarities between different text or question representations. The CLS token was used as the overall 
representation of the input questions. No observable difference was found while using mean pooling strategy.

\vspace*{1mm}
{\bf Evaluation Measures.} We evaluate the performance of the algorithms using the following measures: \\
(i) {\em Mean Average Precision} (MAP) -- computes the mean (over the query set) of average precision using the rank position of relevant QA pairs returned. \\
(ii) {\em Precision-at-Rank-5} (P@5) -- reports the number of relevant answers among the top-$5$ retrieved QA pairs, providing a more practical measure as users 
typically tend to inspect the top few results.

\vspace*{1mm}
{\bf Experimental Setup.} We adopt the setup of~\cite{karan2018}, with $80$-$20$ train-test data split and report the averaged results across five-fold 
cross-validation runs. Further, for supervised model training, FAQ pairs in the train set were provided with relevance annotations with respect to other questions in the form of 
{\em relevance matrix}, i.e., if a FAQ pair $(Q_j, A_j)$ is relevant to $Q_i \in$ FAQ, its corresponding annotation is set to $1$ (i.e., the $(ij)^{th}$ element of the relevance matrix is 
set to $1$), otherwise is considered as $0$.

%
%{\em TI-S2S} uses a sequence-to-sequence 
The Seq2Seq module of {\em TI-S2S} consisting of an LSTM model with $2048$ encoder nodes, {\em concat} Luong Attention mechanism~\cite{luong2015}, and a dropout factor of $0.4$ as regularizer. 
The decoder uses a {\em tanh} activation function optimized for {\em Sparse Categorical Cross-Entropy} loss function. Additionally, for candidate generation, {\em GTI-S2S} 
stacks a Gated Recurrent Unit (GRU) with $1024$ units, a $512$ node fully-connected layer having {\em SoftMax} activation, 
%Rectified Linear Unit (ReLU) activation 
and $0.5$ dropout layer. The models are trained with $32$ batch size over $30$ epochs with TF-IDF threshold $\tau$ set to $0.15$ (refer Section~\ref{ssec:param}), and top-$20$ candidates 
were considered in {\em GTI-S2S}. Publicly available pre-trained Google Word2Vec embeddings were used. For all algorithms, the input questions were pre-processed to remove stopwords and 
were lemmatized.

\subsection{Overall Results}
\label{ssec:result}

The obtained performance results of the competing algorithms are presented in Table~\ref{tab:all_results}(a). The use of TF-IDF to obtain discriminating words characterizing the different 
question groups and learning the transformation of questions to representative keywords via a sequence-to-sequence model provide a proxy to understanding the context, topic and intent of the 
questions. This enables our proposed algorithms {\em TI-S2S} and {\em GTI-S2S} to achieve more than $\textbf{92\%}$ accuracy on both the MAP and P@5 measures. We observe 
that our framework outperforms the existing approaches with nearly ${\bf 13\%}$ improvements in terms of {\em P@5} over RoBERTa, and around $3\%$ better {\em MAP} score over
{\em TSU-BERT}. 

The {\em GTI-S2S} framework depicts a slight increase in performance over {\em TI-S2S}, which can be attributed to the ``guided candidate selections'' from the additional recurrent neural network
based learning module. Although the overall gain is marginal 
for {\em GTI-S2S}, note that this variant provides robustness against sub-optimal parameter settings or minor prediction errors. For example, in 
Table~\ref{tab:all_results}(b), for a sub-optimal TF-IDF threshold setting (e.g., $\tau = 0.05$) the performance of {\em GTI-S2S} is still efficient ($\sim 91\%$ MAP) compared to {\em TI-S2S}. Further, 
as seen in Figure~\ref{fig:train}, {\em GTI-S2S} also performs better in scenarios with limited training data availability.
Thus, the {\em GTI-S2S} provides a robust variant of our algorithm.

\subsection{Parameter Setting}
\label{ssec:param}

The working of our proposed {\em TI-S2S} and {\em GTI-S2S} algorithms depends on the thresholding hyper-parameter $\tau$
on TF-IDF score to extract {\em representative intent target keywords} characterizing the various contexts presents in the 
questions. We study the performance of our algorithm (on MAP and P@5) for different values of $\tau$. Figure~\ref{tab:all_results}(b) shows a ``bell-like'' curve with $\tau = 0.15$ (used in our experiments)
providing the best empirical results.

For interpretability, we list the target intent keywords identified (for training the sequence-to-sequence module) at different values of $\tau$. As seen in Table~\ref{tab:tfidf}, a high threshold value extracts 
only a few {\em representative intent} words which fail to properly model the full context of the QA pairs. For example, in the second row of Table~\ref{tab:tfidf} only 
{\small{\tt ``conversation''}} is identified as the representative keyword (with $\tau \geq 0.4$), completely ignoring the vital context of {\small{\tt ``gmail''}}. On the other hand, a very low value of 
$\tau$ is seen to extract non-informative words which possibly overlaps with other QA groups, diminishing the discriminative power of the framework. Both scenarios are seen to degrade the overall accuracy 
performance of our algorithm. 
% as seen -- thus, $\tau$ needs to be suitably tuned for application domains.
Thus, this parameter captures the domain-specificity and can be suitably tuned for different application domains.

\subsection{Robustness Study}
\label{ssec:robust}

\begin{figure}[t]
\vspace*{-1mm}
	\centering
	\includegraphics[width=0.85\columnwidth]{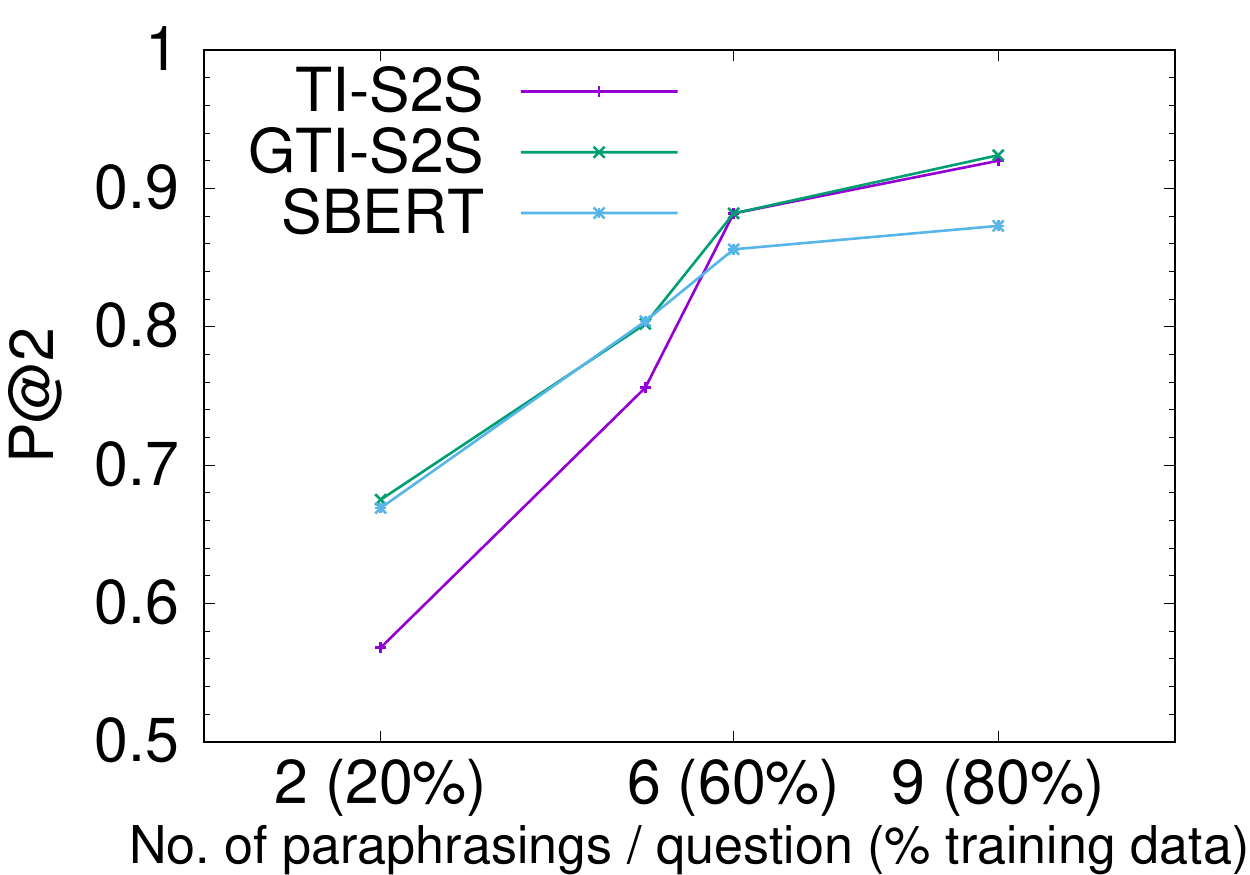}
	\caption{Effect of training size on the performance of the proposed algorithm.}
	\label{fig:train}
\vspace*{-2mm}
\end{figure}

A major challenge for supervised systems is the availability of large annotated training data,
and the ensuing associated costs. The {\em StackExchange} dataset 
used also contains $10$ manual paraphrasing for each original user questions. 
In this regard, we now study the robustness of our approach in presence of limited training data, 
%a major challenge for supervised systems. We 
by varying the number of relevance training questions (paraphrasings with 
same meaning) provided for each question. We compare the performance on P@2 (the smallest training subset has only $2$ variants per QA) with SBERT (demonstrating the best 
P@2 results on the full dataset).
From Figure~\ref{fig:train}, we observe that {\em GTI-S2S} can robustly handle limited supervision scenarios, demonstrating a graceful degradation with performance similar to SBERT 
($\sim 68\%$ accuracy) with only $2$ training examples for each of the $125$ QA threads. However, the accuracy of {\em TI-S2S} is seen to be more affected.
%Hence, the {\em GTI-S2S} algorithm with ``guided candidate generation'' 
Hence, {\em GTI-S2S} with ``guided candidate generation'' can robustly handle applications with limited supervision needs.

Overall, we observe that the proposed {\em TI-S2S} and {\em GTI-S2S} frameworks enable efficient FAQ retrieval by capturing query intent via representative target keywords. Experimental 
results demonstrate the transformation of questions onto a common keyword space provides improved accuracy as well as robustness.

\vspace*{-2mm}
\section{Conclusion}
\label{sec:conc}

We propose a novel FAQ Retrieval system using sequence-to-sequence framework to compute the similarity between user queries and FAQ based on ``predicted representative intent keywords''.
We show how the filter-and-refine approach utilizing TF-IDF scores to obtain the representative keywords of questions act as weak supervision cues for capturing semantic similarities bridging the 
lexical and contextual gap in short span FAQ retrieval systems. Further, we also show that the use of ``candidate identification'' from an additional learning module boosts the performance of our 
framework by enabling early pruning. Experimental results on open-source FAQ dataset demonstrated the efficacy and robustness of our algorithm over existing approaches.

\balance
\bibliographystyle{named}
\bibliography{biblio}

\begin{thebibliography}{}

\bibitem[\protect\citeauthoryear{Aizawa}{2003}]{tfidf}
A.~Aizawa.
\newblock {An Information-Theoretic Perspective of TF-IDF Measures}.
\newblock {\em {Information Processing \& Management}}, 39(1):45--65, 2003.

\bibitem[\protect\citeauthoryear{Bengio \bgroup \em et al.\egroup
  }{2015}]{bengio2015}
S.~Bengio, O.~Vinyals, N.~Jaitly, and N.~Shazeer.
\newblock {Scheduled Sampling for Sequence prediction with Recurrent Neural
  Networks}.
\newblock In {\em {NIPS}}, pages 1171--1179, 2015.

\bibitem[\protect\citeauthoryear{Berger \bgroup \em et al.\egroup
  }{2000}]{berger2000}
A.~Berger, R.~Caruana, D.~Cohn, D.~Freitag, and V.~Mittal.
\newblock {Bridging the Lexical Chasm: Statistical Approaches to
  Answer-finding}.
\newblock In {\em {SIGIR}}, pages 192--199, 2000.

\bibitem[\protect\citeauthoryear{Devlin \bgroup \em et al.\egroup
  }{2019}]{devlin2019}
J.~Devlin, M.~Chang, K.~Lee, and K.~Toutanova.
\newblock {{BERT}: Pre-training of Deep Bidir. Transformers for Lang.
  Understanding}.
\newblock In {\em {NAACL-HLT}}, pages 4171--4186, 2019.

\bibitem[\protect\citeauthoryear{Figueroa}{2017}]{cqa2017}
A.~Figueroa.
\newblock {Automatically Generating Effective Search Queries Directly from
  Community QA Questions for Finding Related Questions}.
\newblock {\em {Expert Systems with Applications}}, 77:11--19, 2017.

\bibitem[\protect\citeauthoryear{Gupta and Carvalho}{2019}]{gupta19}
S.~Gupta and V.~R. Carvalho.
\newblock {FAQ Retrieval Using Attentive Matching}.
\newblock In {\em {SIGIR}}, pages 929--932, 2019.

\bibitem[\protect\citeauthoryear{Jijkoun and de Rijke}{2005}]{jij2005}
V.~Jijkoun and M.~de~Rijke.
\newblock {Retrieving Answers from Frequently Asked Questions Pages on the
  Web}.
\newblock In {\em {CIKM}}, pages 76--83, 2005.

\bibitem[\protect\citeauthoryear{Karan and \v{S}najder}{2018}]{karan2018}
M.~Karan and J.~\v{S}najder.
\newblock {Paraphrase-focused Learning to Rank for Domain-specific FAQ
  Retrieval}.
\newblock {\em {Expert Systems With Applications}}, 91:418--433, 2018.

\bibitem[\protect\citeauthoryear{Kothari \bgroup \em et al.\egroup
  }{2009}]{kothari2009}
G.~Kothari, S.~Negi, T.~A. Faruquie, V.~T. Chakaravarthy, and L.~V.
  Subramaniam.
\newblock {SMS Based Interface for FAQ Retrieval}.
\newblock In {\em {ACL-IJCNLP}}, pages 852--860, 2009.

\bibitem[\protect\citeauthoryear{Kusner \bgroup \em et al.\egroup }{2015}]{wmd}
M.~J. Kusner, Y.~Sun, N.~I. Kolkin, and K.~Q. Weinberger.
\newblock {From Word Embeddings to Document Distances}.
\newblock In {\em {ICML}}, pages 957--966, 2015.

\bibitem[\protect\citeauthoryear{Lee \bgroup \em et al.\egroup
  }{2008}]{lee2008}
J.~T. Lee, S.~B. Kim, Y.~I. Song, and H.~C. Rim.
\newblock {Bridging Lexical Gaps between Queries and Questions on Large Online
  Q\&A Collections with Compact Translation Models}.
\newblock In {\em {EMNLP}}, pages 410--418, 2008.

\bibitem[\protect\citeauthoryear{Liu \bgroup \em et al.\egroup
  }{2019}]{liu2019}
Y.~Liu, M.~Ott, N.~Goyal, J.~Du, M.~Joshi, D.~Chen, O.~Levy, M.~Lewis,
  L.~Zettlemoyer, and V.~Stoyanov.
\newblock {RoBERTa: {A} Robustly Optimized {BERT} Pretraining Approach}.
\newblock {\em {CoRR}}, abs/1907.11692, 2019.

\bibitem[\protect\citeauthoryear{Lommatzsch and Katins}{2019}]{lommatzsch2019}
A.~Lommatzsch and J.~Katins.
\newblock {An Information Retrieval-based Approach for Building Intuitive
  Chatbots for Large Knowledge Bases}.
\newblock In {\em {LWDA}}, pages 343--352, 2019.

\bibitem[\protect\citeauthoryear{Luong \bgroup \em et al.\egroup
  }{2015}]{luong2015}
T.~Luong, I.~Sutskever, Q.~V. Le, O.~Vinyals, and W.~Zaremba.
\newblock {Addressing the Rare Word Problem in Neural Machine Translation}.
\newblock In {\em {ACL-IJCNLP}}, pages 11--19, 2015.

\bibitem[\protect\citeauthoryear{Malik \bgroup \em et al.\egroup
  }{2007}]{malik2007}
R.~Malik, L.~V. Subramaniam, and S.~Kaushik.
\newblock {Automatically Selecting Answer Templates to Respond to Customer
  Emails}.
\newblock In {\em {IJCAI}}, pages 1659--1664, 2007.

\bibitem[\protect\citeauthoryear{Mass \bgroup \em et al.\egroup
  }{2020}]{mass20}
Y.~Mass, B.~Carmeli, H.~Roitman, and D.~Konopnicki.
\newblock {Unsupervised FAQ Retrieval with Question Generation and {BERT}}.
\newblock In {\em {ACL}}, pages 807--812, 2020.

\bibitem[\protect\citeauthoryear{Massaro \bgroup \em et al.\egroup
  }{2018}]{massaro2018}
A.~Massaro, V.~Maritati, and A.~Galiano.
\newblock {Automated Self-learning Chatbot Initially Build as a FAQs Database
  Information Retrieval System}.
\newblock {\em {Informatica (Slovenia)}}, 42(4):515--525, 2018.

\bibitem[\protect\citeauthoryear{Mikolov \bgroup \em et al.\egroup
  }{2013}]{mikolov2013}
T.~Mikolov, I.~Sutskever, K.~Chen, G.~S. Corrado, and J.~Dean.
\newblock {Distributed Represt. of Words and Phrases and their
  Compositionality}.
\newblock In {\em {NIPS}}, pages 3111--3119, 2013.

\bibitem[\protect\citeauthoryear{Nogueira \bgroup \em et al.\egroup
  }{2020}]{nog20}
R.~Nogueira, Z.~Jiang, R.~Pradeep, and J.~Lin.
\newblock {Document Ranking with a Pretrained Sequence-to-Sequence Model}.
\newblock In {\em {Findings of EMNLP}}, pages 708--718, 2020.

\bibitem[\protect\citeauthoryear{Otsuka \bgroup \em et al.\egroup
  }{2018}]{qexp18}
A.~Otsuka, K.~Nishida, K.~Bessho, H.~Asano, and J.~Tomita.
\newblock {Query Expansion with Neural Question-to-Answer Translation for
  FAQ-Based QA}.
\newblock In {\em {WWW}}, page 1063–1068, 2018.

\bibitem[\protect\citeauthoryear{Reimers and Gurevych}{2019}]{reimers2019}
N.~Reimers and I.~Gurevych.
\newblock {Sentence-BERT: Sentence Embeddings using Siamese BERT-Networks}.
\newblock In {\em {EMNLP}}, pages 3982--3992, 2019.

\bibitem[\protect\citeauthoryear{Sakata \bgroup \em et al.\egroup
  }{2019}]{sakata2019}
W.~Sakata, T.~Shibata, R.~Tanaka, and S.~Kurohashi.
\newblock {FAQ Retrieval using Query-Question Simi. and BERT-Based QA
  Relevance}.
\newblock In {\em {SIGIR}}, pages 1113--1116, 2019.

\bibitem[\protect\citeauthoryear{Santos \bgroup \em et al.\egroup
  }{2020}]{por2020}
J.~Santos, L.~Duarte, J.~Ferreira, A.~Alves, and H.~G. Oliveira.
\newblock {Developing Amaia: A Conversational Agent for Helping Portuguese
  Entrepreneurs—An Extensive Exploration of Question-Matching Approaches for
  Portuguese}.
\newblock {\em {Information}}, 11(9):428, 2020.

\bibitem[\protect\citeauthoryear{Sneiders}{2010}]{sneider2010}
E.~Sneiders.
\newblock {Automated Email Answering by Text Pattern Matching}.
\newblock In {\em {ICNLP}}, pages 381--392, 2010.

\bibitem[\protect\citeauthoryear{Surdeanu \bgroup \em et al.\egroup
  }{2011}]{nonfact2011}
M.~Surdeanu, M.~Ciaramita, and H.~Zaragoza.
\newblock {Learning to Rank Answers to Non-factoid Questions from Web
  Collections}.
\newblock {\em {Comp. Ling.}}, 37(2):351--383, 2011.

\bibitem[\protect\citeauthoryear{Sutskever \bgroup \em et al.\egroup
  }{2014}]{seq2seq}
I.~Sutskever, O.~Vinyals, and Q.~V. Le.
\newblock {Sequence to Sequence Learning with Neural Networks}.
\newblock In {\em {NIPS}}, pages 3104--3112, 2014.

\bibitem[\protect\citeauthoryear{Xie \bgroup \em et al.\egroup
  }{2019}]{xie2019}
R.~Xie, Y.~Lu, F.~Lin, and L.~Lin.
\newblock {FAQ-based Question Answering via Knowledge Anchors}.
\newblock {\em CoRR}, abs/1911.05930, 2019.

\bibitem[\protect\citeauthoryear{Yan \bgroup \em et al.\egroup
  }{2016}]{yan2016}
Z.~Yan, N.~Duan, J.~Bao, P.~Chen, M.~Zhou, Z.~Li, and J.~Zhou.
\newblock {DocChat: An Information Retrieval Approach for Chatbot Engines using
  Unstructured Docs}.
\newblock In {\em {ACL}}, pages 516--525, 2016.

\end{thebibliography}

\end{document}